\documentclass[prb,showpacs,amsmath,amssymb,twocolumn]{revtex4}

\usepackage{graphicx}
\usepackage{epsfig}
\usepackage{dcolumn}
\usepackage{bm}

\begin{document}

\title{Quantum Dots and Etch-Induced Depletion of a Silicon 2DEG}

\author{L.~J.~Klein$^1$, K.~L.~M. Lewis$^1$, K.~A.~Slinker$^1$, Srijit~Goswami$^1$, 
D.~W.~van~der~Weide$^2$, R.~H.~Blick$^2$, P.~M.~Mooney$^3$, J.~O.~Chu$^3$,
S.~N.~Coppersmith$^1$, Mark~Friesen$^1$, and Mark~A.~Eriksson$^1$}
\affiliation{
$^1$Department of Physics, University of Wisconsin-Madison, Madison, WI 53706-1390, \\
$^2$Electrical and Computer Engineering Department, University of Wisconsin-Madison, 
Madison, WI 53706-1691, \\
$^3$IBM Research Division, T.~J.~Watson Research Center, NY 10598}

\begin{abstract}
The controlled depletion of electrons in semiconductors is the basis for numerous devices.
Reactive-ion etching provides an effective technique for fabricating both classical and
quantum devices. 
However, Fermi level pinning can occur, and must be carefully
considered in the development of small devices, such as
quantum dots.  Because of depletion, the electrical size of the device is reduced in
comparison with its physical dimension.  To investigate this issue, we fabricate several 
types of devices 
in silicon-germanium heterostructures using two different etches, CF$_4$ and SF$_6$.  
We estimate the depletion width associated with each etch by two methods:
(i) conductance measurements in etched wires of decreasing thickness (to determine the 
onset of 
depletion), (ii) capacitance measurements of quantum dots
(to estimate the size of the active region).  We find that the 
SF$_6$ etch causes a much smaller depletion width, making it more suitable for 
device fabrication.  
\end{abstract}

\maketitle

\section{Introduction}
Recently, there has been a renewed interest in modulation doped, silicon-germanium quantum wells.  Modulation doped field-effect transistors (MODFETs) are potentially attractive for communications applications because of their low noise, low cost, high speed, and compatibility with CMOS logic \cite{koester}.  They are also the basis for the high-mobility devices used in quantum dot quantum computing \cite{loss98,vrijen00,friesen03,friesen04} and spintronics, where quantum coherence plays a key role.  Indeed, it is expected that silicon will exhibit the most desirable coherence properties of any semiconductor (except perhaps carbon) because of its weak spin-orbit coupling and the availability of spin-zero nuclear isotopes \cite{kane98,yablonovitch}.  In the context of quantum computing, the goal is to fabricate coupled quantum dots containing individual electrons whose spins act as qubits \cite{friesen03}.  We have made recent progress towards this goal, overcoming several materials and fabrication challenges \cite{QIP}.  In this paper, we report on Coulomb blockade and single electron tunneling experiments in which the number of electrons in a silicon quantum dot can be held constant for up to 11 hours.  This fulfills an important milestone towards qubit fabrication.

During the 1990's, advances in materials science and growth techniques led to 
the development of high quality silicon-germanium quantum wells containing 
two-dimensional electron gases (2DEGs) with low-temperature electron mobilities of order 
$6\times 10^5\, {\rm cm}^2 /$Vs \cite{schaeffler92,ismael95a,sugii98,okamoto04,lai04}.  
Further progress was hampered by difficulties with leakage currents and parallel conduction paths \cite{kanjanachuchai98,notargiacomo03}, and unavoidable defects associated with the growth of strained heterostructures.  However, spurred on by quantum computing, and new technologies like atomic layer oxide deposition, some important technological hurdles have been overcome.  Reactive ion etching has emerged as an important fabrication tool and a viable alternative to top-gating \cite{klein04}.  In this procedure, devices like quantum dots are formed by physically carving the 2DEG.  External side gates for electrostatic
control can be fabricated in nearby regions of 2DEG by the same method.  

In this paper, we present evidence for Coulomb blockade in etched quantum dots and we investigate the depletion of the 2DEG near etched surfaces.  Etching damages 
the crystalline structure, potentially introducing dangling bonds and trapped charge.  
The resulting surface states cause local pinning of the Fermi level near midgap and, 
consequently, local modulation of the 
conduction band energy.  Thus, an electronically depleted region forms near the etch trench.  In some cases, edge depletion can present a challenge for fabricating small devices.  However, there is also a benefit for quantum devices:  because of depletion, the active 2DEG may be physically insulated from defects and decoherence-causing centers at the etch boundary. 

\begin{figure}
\centerline{\includegraphics[width=2.9in]{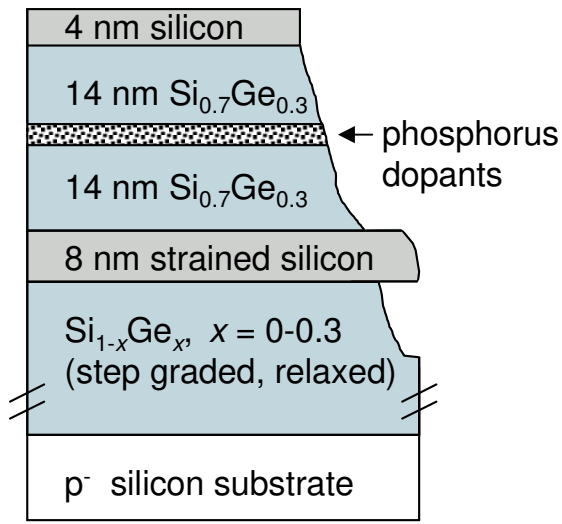}}
\caption{ 
Cartoon sketch of the heterostructures used in this work.  The inhomogeneous region shows 
the result of reactive ion etching.  Etching 
proceeds more slowly for the silicon active layer, causing a ``silicon ledge."
\label{fig:profile}}
\end{figure}

Here, we estimate the depletion width of etched samples by two different methods.  
First, we use the etch process to fabricate arrays of wires with different widths.  Through conductance measurements, we observe the onset of depletion in the narrowest wires and a linear dependence of conductance {\it vs.}\ wire width for larger wires.  Second, we determine the capacitance of etched quantum dots.  Through realistic device simulations, we estimate the electrical size of the dots, and the corresponding depletion widths.  The estimates for wires and dots are in reasonable agreement.  
However, the two etches used here, CF$_4$ and
SF$_6$, cause very different depletion widths:  200-250~nm for CF$_4$, and a vanishing width for SF$_6$.  We attribute these differences to the number or nature
of the surface states.

\section{Growth and Fabrication}
A Si/SiGe heterostructure was grown by ultra-high vacuum chemical vapor deposition (UHVCVD) \cite{ismael95a,klein04}.  The 2DEG sits near the top of an 8~nm strained silicon quantum well, grown on a strain-relaxed Si$_{0.7}$Ge$_{0.3}$ buffer layer.  The 2DEG is topped with a 14~nm Si$_{0.7}$Ge$_{0.3}$ spacer layer, and a 14~nm Si$_{0.7}$Ge$_{0.3}$ supply layer with phosphorus donors.  The final structure is capped with 35~\AA~of silicon, as shown
in Fig.~\ref{fig:profile}. 

\begin{figure}
\centerline{\includegraphics[width=2.5in]{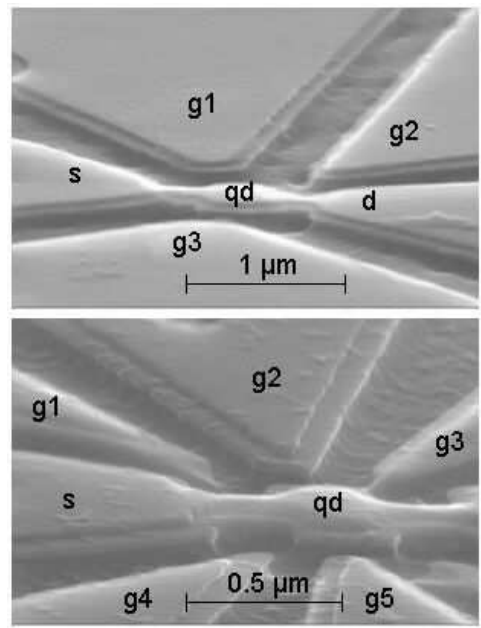}}
\caption{ 
Scanning electron micrographs of 3-gate and 6-gate quantum dots studied in this paper.
The etched structures include source (s) and drain (d) leads, a quantum dot (qd), and  
several side-gates (g1-g6).  (In the lower dot, gate 6 and the drain lie outside the 
viewing area.)
\label{fig:SEMs}}
\end{figure}

Ohmic contacts were formed with the 2DEG by evaporating Au with 1\% Sb and sintering at $400\, ^\circ$C for 10 minutes.  For all experiments described below, the contacts are made prior to reactive-ion etching.  Using a hall-bar geometry, the electron density of the 2DEG is found to be $4\times 10^{11}$~cm$^{-2}$, with a mobility of $40,000$~cm$^2/$Vs at 2~K.   

The devices studied here were fabricated by electron beam lithography and subsequent CF$_4$ or SF$_6$ reactive-ion etching.  Extensive studies of this process have shown that for
most ion species (including those used here), the etch rate of SiGe increases with 
germanium content \cite{zhang93}.  However, other factors, like rf power, gas pressure,  plasma creation processes, and preferential sputtering of silicon, also affect the etch rate.
As a consequence, it is quite difficult to etch uniformly through 
compositionally varying heterostructures.

For the devices used in this work, the heterostructures were etched to a depth of 
approximately 120~nm -- far deeper than the 2DEG layer.   
Fig.~\ref{fig:SEMs} shows scanning electron microscope (SEM) images of our 3-gate dot and
our 6-gate dot.  We observe some rounding near the bottom of the trenches, and a prominent silicon ledge of width $\sim 100$~nm, due to 
the preferential etching of germanium (see Fig.~\ref{fig:profile}).  
The undercutting below the silicon ledge is slightly non-uniform.   

In the following, we measure the depletion widths from the inner perimeter of the 
silicon ledge.   Inside this ledge, the silicon layer remains mainly intact.  However,
the contained 2DEG is depleted, due to strain effects and the close proximity to 
surface trap states.  In the unetched, ``bulk" regions of our devices, we observe differing
depletion widths, depending on the etch.

\section{Etched Wires} \label{sec:wires}
\begin{figure}
\centerline{\includegraphics[width=2.5in]{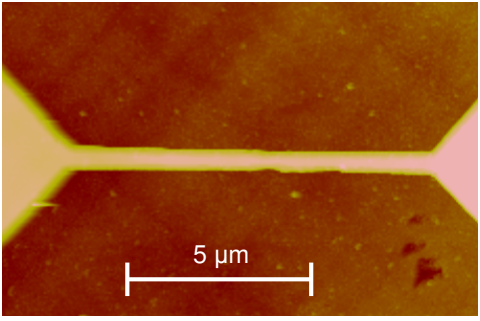}}
\caption{ 
Atomic force microscope image of a fabricated wire.  The darker regions are etched, leaving a narrow channel between the leads.
\label{fig:wireAFM}}
\end{figure}

Nine wires were patterned on two different chips using the CF$_4$ reactive-ion etch.  The wires had different widths, ranging from 200~nm to 1350~nm, and all the devices were  electrically isolated from one another.  A typical device is shown in Fig.~\ref{fig:wireAFM}.

As discussed above, the wires have an electrical width smaller than their physical width, due to the depletion of the 2DEG near the etch boundaries.  This effect has been studied previously in patterned wires \cite{holzmann95,vanveen99,giovine01}.  Here, we investigate edge depletion through transport measurements.  The conductance of the wires should be proportional to their electrical width.  For progressively narrow wires, the 2DEG will eventually cease to conduct.  The onset of zero conductance therefore indicates full depletion, and provides an estimate of the depletion width.  We assume the depletion width is the same for all the wires on a given chip, due to identical fabrication conditions.

\begin{figure}[b]
\centerline{\includegraphics[width=3in]{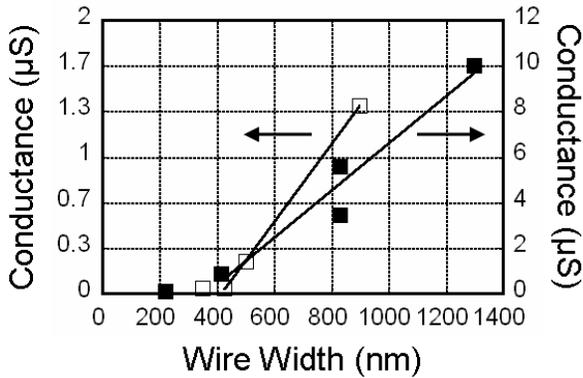}}
\caption{ 
Conductance vs.\ wire width, for nine wires on two different chips.  The different
symbols correspond to different chips.  The lines show the result of fitting to 
the non-zero data.  The $x$-intercepts give an estimate of ($2\times$) the depletion width.  
\label{fig:wireconductance}}
\end{figure}

In Fig.~\ref{fig:wireconductance}, the two-point conductance is plotted for each of the wires, as a function of their physical width.  Two-point measurements were used instead of four-point measurements in order to maximize the number of wires on a given chip for a given cool-down.  Each data point represents an individual wire.  The two chips are identified by different symbols.  All the measurements are made at 2~K.  

As expected, the wire conductance was found to vary approximately linearly with the physical width.  A line fit through the higher data points intersects the $x$-axis at about 400~nm for both chips, suggesting a depletion width of about $200$~nm on either side of the wires.    

\section{Quantum Dots}
Several quantum dots were fabricated by reactive ion etching.  Here, we report on the two dots shown in Fig.~\ref{fig:SEMs}.  Although similar etching 
procedures were applied in the two cases, the 3-gate dot used the CF$_4$ etch, while the 6-gate dot used the SF$_6$ etch.  All the components of the devices (dot, leads, and gates) were etched from the same 2DEG, with no additional metal top-gates.  
Because of the considerable etch depth (120~nm), we could detect no current flow between the side gates and the dots.  However, in other experiments, not reported here, small leakage currents were observed for shallower etch depths ($<100$~nm).  We conjecture that dislocation arms running underneath shallow etch trenches may connect the gates to the dot in such devices.  Here, all side gates were contacted ohmically and grounded, unless otherwise noted.

The source and drain leads, together with the quantum dot, are formed from a single 
structure.  However, the necks between the dot and the source-drain leads are designed to give transport resistances slightly larger than the resistance quantum 
$e^2/h\simeq 25\,{\rm k}\Omega$.  Thus, the necks form tunnel barriers, 
even when the side-gates are grounded.  
By adjusting the voltages on different side-gates,
we can tune the number of electrons on the dot and the tunnel couplings to the source and 
drain leads.  Additional details are provided in Ref.~\onlinecite{klein04}.   

\begin{figure}
\centerline{\includegraphics[width=2.8in]{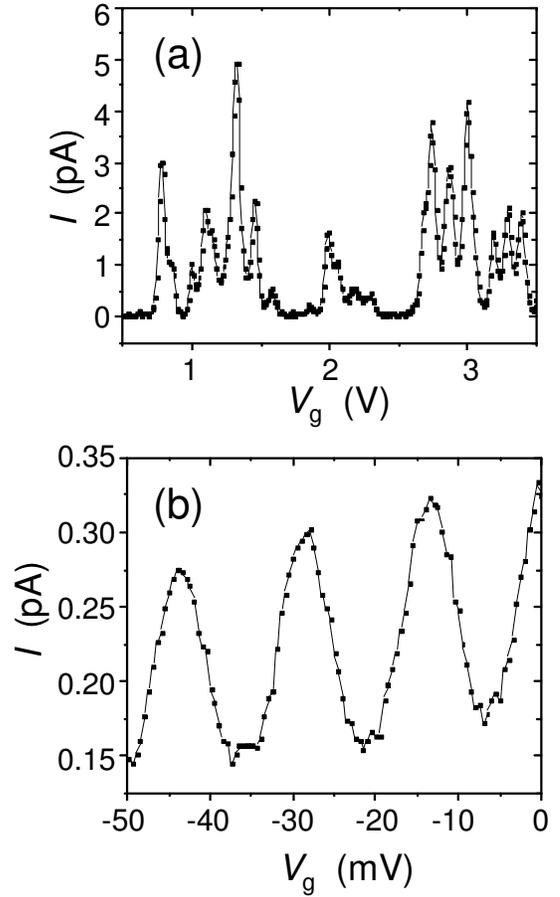}}
\caption{
Coulomb blockade current peaks as a function of gate voltage for 
(a) the 3-gate  dot, and (b) the 6-gate dot.  
\label{fig:CB}}
\end{figure}

Several types of transport measurements are performed on the dots.  First, we measure
the current through the quantum dot while
varying the voltage of a side-gate $V_g$, but holding the bias voltage between
the source and drain constant.  As shown in Fig.~\ref{fig:CB},
we observe periodic oscillations of the transport current through the quantum dots.
The peaks indicate charge quantization in the dots,
with successive peaks corresponding to an increment of one electron \cite{kouwenhoven}.   

The spacing between consecutive peaks $\Delta V_g$ is governed primarily by the plunger gate capacitance, $C_g=e/\Delta V_g$, which determines the energy of the dot relative to the leads.  For the 3-gate dot, we observe gate capacitances between 0.7 and 1.3 aF 
\cite{klein04}.  Note that, because of the larger separations (and weaker couplings) between 
dots and gates in side-gate devices, 
various combinations of gates can be used for plunging, giving slightly 
different capacitances.  For the 6-gate dot, we tie together two of the six 
gates ($g1$ and $g5$ in Fig.~\ref{fig:SEMs}) to produce the gate
voltage $V_g$, while grounding the other gates.  The resulting data in 
Fig.~\ref{fig:CB}(b) correspond to a gate capacitance of 11~aF.  Thus, the 6-gate dot
has a much larger gate capacitance than the 3-gate dot,
signaling a smaller depletion width in the former.

\begin{figure}
\centerline{\includegraphics[width=3in]{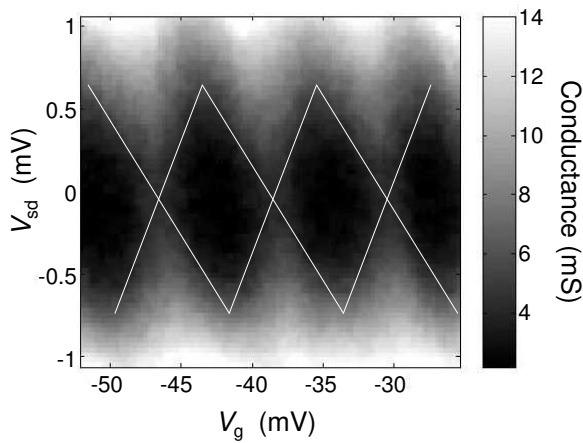}}
\caption{ 
Stability plot for the 6-gate dot, as function of gate and source-drain voltages.
Measurements were performed at 300~mK, and show no switching events over the 11~hour
measurement window.  Diamonds are drawn as a guide to the eye.
\label{fig:stability}}
\end{figure}

A stability plot for the 6-gate dot is shown in Fig.~\ref{fig:stability}.  The data were 
taken over an 11 hour window, in which no charge switching events occurred.  This represents a
significant improvement over the 3-gate dot, as reported in Ref.~\onlinecite{klein04}.  
In part, we attribute the enhancement to the etching process:  the
SF$_6$ etch appears to generate a lower density of surface states, and a corresponding
lower frequency of charge switching.  However, the measurement was also performed in a 
3.6~T magnetic field, known to reduce charge noise.  

The shape of the Coulomb diamonds in Fig.~\ref{fig:stability} 
provides further information about the dot capacitance 
$C_\Sigma$.  In the conventional theory of Coulomb blockade \cite{devoret}, 
transport phenomena are explained in terms of the chemical 
potentials of the $N$-electron dot, and the left and right leads, $\mu_{\rm dot}(N)$, 
$\mu_L$, and $\mu_R$, respectively.
At low temperatures, transport across the dot occurs only when
$\mu_L>\mu{\rm dot}(N) \simeq \mu_{\rm dot}(N+1)>\mu_R$.
Such considerations allow calculations of the shape of the diamonds.  The resulting height 
(from tip to tip) is given by $2e/C_\Sigma$, from which we determine $C_\Sigma$.  Thus, in
Ref.~\onlinecite{klein04}, we determined $e^2/C_\Sigma = 3.2$~meV for the 3-gate dot.
For the 6-gate dot, the analysis has large error because of the broad conductance
peaks in Fig.~\ref{fig:CB}(b).  We find $C_\Sigma \simeq = 230$~aF, 
with a charging energy of 0.70~meV.

\begin{figure}
\centerline{\includegraphics[width=2.7in]{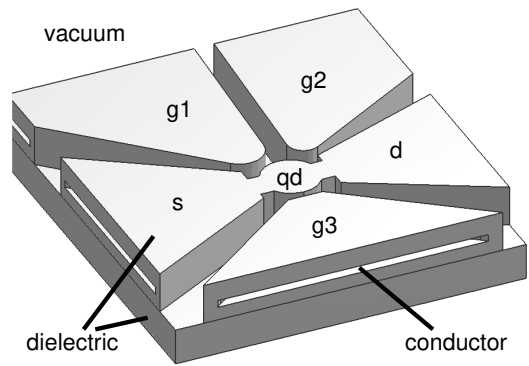}}
\caption{Simulation geometry for the 3-gate dot.
\label{fig:simulationgeom}}
\end{figure}

We can use these capacitance values to estimate the diameter of the electrically active regions of the quantum dots.  Treating the dot as a conducting disk surrounded by an infinite dielectric obtains the conductance formula
\begin{equation}
C_\Sigma = 4 \epsilon D, \label{eq:CSigma}
\end{equation}
where $\epsilon = 11.4\epsilon_0$ is the low-temperature dielectric constant of silicon and 
$D$ is the diameter of the disk.  (Note that in our calculations, we assume 
materials parameters consistent with silicon, rather than Si$_{0.7}$Ge$_{0.3}$.  This is
convenient because (i) the dielectric constants of the two materials differ 
by only 6\% at room temperature \cite{ahuja03}, (ii) low temperature parameters for SiGe 
are not well known.)  Eq.~(\ref{eq:CSigma}) is clearly an approximation.  It does not
incorporate
the complex structure of the real device.  In particular, it does not
take into account the bending of the electric field lines at
the vacuum interface or the metallic leads.  For side-gate devices, the interface
has a stronger effect, so that Eq.~(\ref{eq:CSigma}) over-estimates $C_\Sigma$ for
a given $D$.  In Sec.~\ref{sec:sims}, we treat this problem more carefully using numerical 
techniques.
 
From Eq.~(\ref{eq:CSigma}), we can back-out an estimate for the electrical width of the 
dot (as opposed to the actual physical width), using
our estimates for the dot capacitance.  For the 3-gate dot, we find an electrical diameter of
124~nm, compared to a physical width of 720~nm (measured from the silicon ledge).  For the 6-gate dot, we find an electrical width of 570~nm.  Within our error bars, this is
the same as its physical width, given by 560~nm.  Since the approximations
leading to Eq.~(\ref{eq:CSigma}) under-estimate the size of the dot, 
the depletion width in the 6-gate dot must be vanishingly small. 
Thus, the capacitance measurements indicate 
a much larger depletion width for the 3-gate dot.

\begin{figure}
\centerline{\includegraphics[width=2.5in]{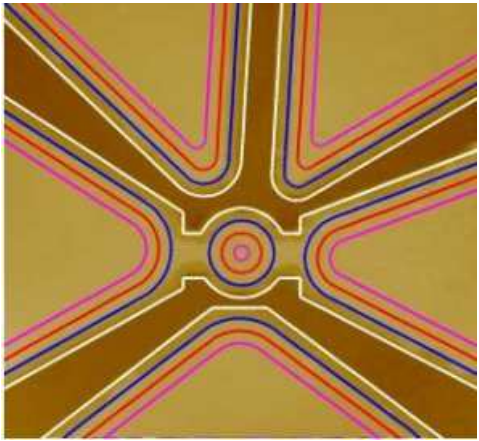}}
\caption{
Top-view of the simulation geometry (Fig.~\ref{fig:simulationgeom}),
showing the etch boundaries (white lines), and the boundaries
of the buried 2DEG (colored lines).  Three depeletion boundaries are shown, corresponding
to three different depletion widths.
The curves overlay an AFM image of the 3-gate
dot.  The imaged area is 4~$\mu$m~$\times$~4~$\mu$m. 
\label{fig:simulationAFM}}
\end{figure}

Based on our measurement of the electron sheet density, far from the etched
boundaries, and our estimates of the dot diameters, we can determine the number of 
electrons in our dots.  Using Eq.~(\ref{eq:CSigma}), we find 50 electrons for the 
3-gate dot.  (A better estimate for the dot diameter, in the following section,
gives 170 electrons.)  For the 6-gate
dot, we estimate 1000 electrons.  We point out that in spite of this large number,
single electron charging effects are still prominent, and the operation is very stable.

\section{Modeling of the Quantum Dot} \label{sec:sims}
We can improve on the estimate of Eq.~(\ref{eq:CSigma}) by performing more accurate
numerical modeling of the physical device.
We specifically consider the 3-gate dot of Fig.~\ref{fig:SEMs}.  

The three-dimensional model geometry used in our simulations is shown in 
Fig.~\ref{fig:simulationgeom}.  The solid regions correspond to dielectric material, which we model with silicon materials parameters, as discussed above.  The exclusions in the solid 
are metallic, corresponding to undepleted 2DEG regions in the gates and leads.  These are 
properly modeled by fixed-voltage boundary conditions.  

Since the depletion width is not known initially, we perform our simulations for many different 2DEG boundaries.  A top-view of several of the
boundaries is shown in Fig.~\ref{fig:simulationAFM}.  The curves overlay an AFM image of the device.  The depletion boundaries are taken to be equidistant from the physical boundaries.  This separation corresponds to the depletion width.  The necks between the dot and the source-drain leads are assumed to be fully depleted, as consistent with experimental evidence. 

The capacitance matrix, relating charges to voltages for a system of conductors, is given by \cite{jackson} $Q_i=\sum_{j=1}^nC_{ij}V_j$.  The conventional capacitances correspond to the diagonal coefficients $C_{ii}$.  These are computed in our simulation by setting the voltages on all the gates to zero, except for the conductor of interest.  Thus, the calculation of $C_\Sigma$ proceeds by setting the voltage of the quantum dot to 1~V, and the other gates 
to 0~V.  The resulting charge on the quantum dot is equal to its capacitance 
in units of Farads (F).  

In this way, a finite element analysis \cite{femlab} obtains the capacitance of the dot as a function of its size.  For the experimental value $C_\Sigma =50$~aF, our simulations give 
a corresponding dot diameter of 233~nm, and a depletion width of 244~nm.  This dot size is nearly two times larger than the estimate of 
Eq.~(\ref{eq:CSigma}).  Thus, the numerical result more closely matches the estimate
obtained in Sec.~\ref{sec:wires}, and confirms our expectation that Eq.~(\ref{eq:CSigma}) 
should under-estimate the true electrical diameter.  

\section{Discussion} \label{sec:discussion}
We have seen that devices fabricated with the CF$_4$ and SF$_6$ etches exhibit some important differences.  For the CF$_4$ etch, we observe a significant depletion
width, while for the SF$_6$ etch, the electrical width of the dot
is nearly the same as its physical width.  We believe these differences
can be attributed to a change in the number or the nature of surface charge states 
caused by the different etching procedures.  Yet, there is a contradiction, which
is most striking for the SF$_6$ dot.  Although the dot itself remains undepleted, this
cannot be true for the adjacent tunnel barriers.  How can we understand this apparently 
non-uniform depletion?

There are several competing effects which contribute to the observed behavior.  The
dominant effect is electrostatic. 
The surface states attract charge from the supply layer, forming a band that
pins the Fermi level near midgap \cite{sze}.  The redistribution of charge induces
electric fields, which bend the conduction band locally.  
The depletion of the 2DEG near the silicon ledge therefore
depends on the number of filled traps, and their position in the band gap.  
In the vicinity of the quantum dot, the density of trapped charge at the etch boundary
is high, due to the convergence of side-gates and other structures.  
Thus, there should be a local enhancement of the depletion width, due to
electrostatic effects.  By this argument, we might expect a larger depletion 
width in a quantum dot than a wire.  Indeed, this seems to be the case for
our experiments.  

\begin{figure}
\centerline{\includegraphics[width=2.5in]{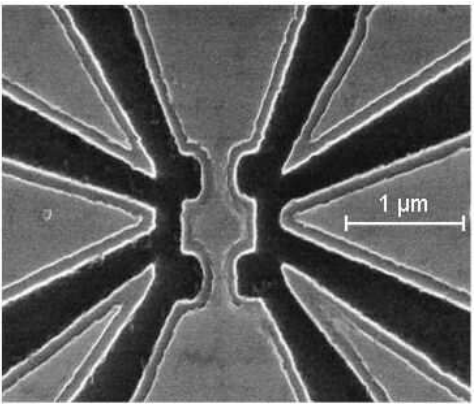}}
\caption{ 
Atomic Force microscope image of the 6-gate dot in Fig.~\ref{fig:SEMs}.  
The silicon ledge is observed as a $\sim 100$~nm rim around each gate. 
\label{fig:dotAFM}}
\end{figure}

Because the walls of the tunnel barriers are so closely spaced, 
we also expect this part of the device to exhibit enhanced depletion.
To model the electrostatics of the surface states, we perform a simulation of the 
6-gate dot, introducing a uniform charge 
density on both the top and etched surfaces.  (The boundary of the etched structures is
taken to be the inner perimeter of the silicon ledge, rather than the outer perimeter --
see Fig.~\ref{fig:dotAFM}.)  For simplicity, we do not include screening by the 2DEG.
Such a model is incomplete, but it provides insight into the effects of surface charge.
The computed electrostatic potential gives a good approximation of the locally varying 
conduction
band.  In a real device, depletion occurs when the band energy lies above the Fermi level.
In our model, depletion occurs when the potential is larger than a particular equipotential
line.  

\begin{figure}
\centerline{\includegraphics[width=2.25in]{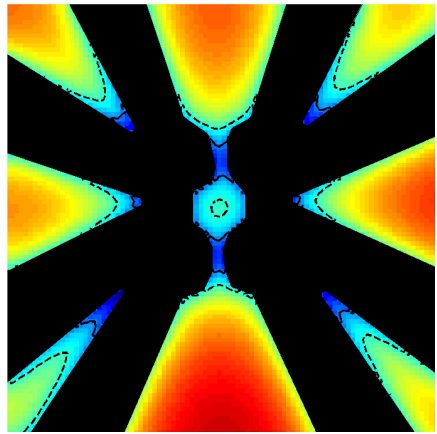}}
\caption{Color scale results for the electrostatic potential 
of the 6-gate dot (arbitrary units), with etch boundaries at the inner silicon ledge.
(See Fig.~\ref{fig:dotAFM}.)  
The potential is evaluated at the 2DEG layer.  Two equipotential lines are shown.  
The solid curve shows a depletion boundary consistent with capacitance 
measurements in the 6-gate dot.
\label{fig:deplete}}
\end{figure}

Two possible depletion boundaries are shown in Fig.~\ref{fig:deplete}.
The solid line describes a situation most consistent 
with the SF$_6$ etch and the 6-gate quantum dot.  Here, the dot and the leads show almost
no depletion, while the tunnel barriers are fully depleted.  The tips of the
side-gates are also partially depleted, as most evident in the narrower gates.  
The dashed line shows a different equipotential with much greater depletion.  
Because Poisson's equation is linear, the same outcome could be obtained with
the former equipotential criterion, but a larger surface charge density.  (In contrast,
the self-consistent simulations discussed below are non-linear.)
The dashed line shows significant depletion of the
quantum dot, as well as in the source, the drain, and the side-gates.  The overall picture is consistent with our measurements in the 3-gate dot.
We conclude that devices using the CF$_4$ etch are more strongly affected 
by the electrostatics of surface states.

In addition to electrostatic effects, quantum kinetic energy 
can also contribute to the depletion of a narrow tunnel barrier.
Typically, such effects becomes noticeable in devices smaller 
than $\sim 100$~nm.  In Fig.~\ref{fig:dotAFM}, we see the minimum barrier width
(not including the silicon ledge) can be as small as 80~nm.  Treating the 
barrier as a square well gives the transverse confinement energy 
$\hbar^2\pi^2/2m^*L_b^2 =0.3$~meV, where $L_b$ is the barrier width.  
If the effective width of the barrier is further 
reduced by the electrostatic potential from the surface charges, the  
confinement energy can easily reach several meV.  Since this is a characteristic 
energy scale for quantum dot devices, it is reasonable to assume that confinement 
effects may also play a role in the depletion of narrow channels. 

To facilitate the development of etched devices, 
it would be desirable to perform simulations that include both 
electrostatic and quantum mechanical effects, while allowing charge 
to move between the trap states, the supply layer, and the 2DEG, to satisfy
electrochemical equilibrium.  A goal would be
to determine the charge profile and the depletion boundary 
self-consistently.  A recent theory by Fogler has made steps in this direction
by determining the depletion boundary self-consistently
\cite{fogler}.  Charge redistribution and quantum effects are not yet included
in the theory.

\section{Conclusions}
We have obtained experimental estimates for the 2DEG depletion width in etched silicon heterostructures by (i) a conductance technique involving etched wires and (ii) a capacitance technique involving etched quantum dots.  For the CF$_4$ etch, the two 
estimates give 200~nm and 244~nm, respectively.  For the SF$_6$ etch, we 
observe a small or vanishing depletion width by the capacitance technique.  
Thus, SF$_6$ appears especially useful for fabricating small devices.  

By means of simulations, we have shown that the depletion width is 
locally varying in etched devices, as consistent with the experimental observations.
The shape of the depletion boundary depends many factors, presenting a challenge 
for device design.  
However, we believe that theoretical and numerical
tools should be available in the near future.

The proximity between gates and dots forms an important control issue, since it determines the capacitive coupling of the side-gates.  For the devices studied here, the minimum separation between undepleted regions in neighboring dots or side-gates is about 
200~nm in SF$_6$-etched devices, reflecting the width of the silicon ledge.  We believe this level of control will enable the development of many types of silicon devices for experiments and applications.  In the context of quantum computing, this work forms a promising step towards silicon qubits.  Ultimately, such devices will require a fabrication resolution as small as 
50-150~nm \cite{friesen03}.  We anticipate that improvements in theoretical design tools, 
and etching and top-gating techniques, will enable us to achieve this goal.

\section*{Acknowledgment}
The authors would like to acknowledge Mariana Lazar for providing her imaging code.
This work was supported by NSA and ARDA under ARO contract  
number W911NF-04-1-0389, and by the National Science Foundation through the ITR 
program (DMR-0325634), the QuBIC program (EIA-0130400), and the MRSEC program
(DMR-0079983).


\end{document}